\documentclass[superscriptaddress, reprint]{revtex4-2}
\usepackage{multirow}
\usepackage{hyperref}
\usepackage[version=4]{mhchem} 
\usepackage{graphicx}
\usepackage{chemmacros}
\usepackage{gensymb}
\usepackage{multirow}
\usepackage{dcolumn}
\usepackage{amssymb}

\begin{document}    

\title{Competing charge and magnetic order in the candidate centrosymmetric skyrmion host \ce{EuGa2Al2}}

\author{A. M. Vibhakar}
\affiliation{Diamond Light Source Ltd,  Harwell Science and Innovation Campus, Didcot, Oxfordshire, OX11 0DE, United Kingdom}
\author{D. D. Khalyavin}
\affiliation{ISIS facility, Rutherford Appleton Laboratory-STFC, Chilton, Didcot, OX11 0QX, United Kingdom}
\author{J. M. Moya}
\affiliation{Applied Physics Graduate Program, Smalley-Curl Institute, Rice University, Houston, Texas 77005, USA}
\affiliation{Department of Physics and Astronomy, Rice University, Houston, Texas, 77005 USA}
\author{P. Manuel}
\affiliation{ISIS facility, Rutherford Appleton Laboratory-STFC, Chilton, Didcot, OX11 0QX, United Kingdom}
\author{F. Orlandi}
\affiliation{ISIS facility, Rutherford Appleton Laboratory-STFC, Chilton, Didcot, OX11 0QX, United Kingdom}
\author{S. Lei}
\affiliation{Applied Physics Graduate Program, Smalley-Curl Institute, Rice University, Houston, Texas 77005, USA}
\affiliation{Department of Physics and Astronomy, Rice University, Houston, Texas, 77005 USA}
\author{E. Morosan}
\affiliation{Applied Physics Graduate Program, Smalley-Curl Institute, Rice University, Houston, Texas 77005, USA}
\affiliation{Department of Physics and Astronomy, Rice University, Houston, Texas, 77005 USA}
\author{A. Bombardi}
\affiliation{Diamond Light Source Ltd,  Harwell Science and Innovation Campus, Didcot, Oxfordshire, OX11 0DE, United Kingdom}
\affiliation{Clarendon Laboratory, Department of Physics, University of Oxford, Parks Road, Oxford OX1 3PU, UK}

\date{\today}

\begin{abstract}
\ce{Eu(Ga_{1-x}Al_x)4} are centrosymmetric systems that have recently been identified as candidates to stabilise topologically non-trivial magnetic phases, such as skyrmion lattices. In this Letter, we present a high-resolution resonant x-ray and neutron scattering study on \ce{EuGa2Al2} that provides new details of the complex coupling between the electronic ordering phenomena. Our results unambiguously demonstrate that the system orders to form a spin density wave with moments aligned perpendicular to the direction of the propagation vector, and upon further cooling, a cycloid with moments in the $ab$ plane, in contrast to what has been reported in the literature. We show that concomitant with the onset of the spin density wave is the suppression of the charge order, indicative of a coupling between the localised 4$f$ electrons and itinerant electron density. Furthermore we demonstrate that the charge density wave order breaks the four-fold symmetry present in the $I4/mmm$ crystal structure, thus declassifying these systems as square-net magnets. 
\end{abstract}

\maketitle

\section{Introduction}

The formation of skyrmion lattices (skLs) in materials with centrosymmetric symmetry has prompted an interest in the scientific community to pinpoint the mechanisms stabilising skyrmions formation in the absence of Dzyaloshinskii-Moriya exchange (DM) interactions \cite{2020Ishiwata, 2012Okubo, 2015Leonov, 2016Batista, 2016Zeng, 2017Ozawa, 2021Yambe}. The absence of dominant sources of anisotropy and the presence of weakly competing exchange interactions is common to many of the centrosymmetric skyrmion hosts, however the exact mechanism that leads to their formation is not yet understood. Thus far several mechanisms have been proposed, such as geometrical frustration \cite{2012Okubo, 2015Leonov}, Ruderman-Kittel-Kasuya-Yosida (RKKY) interactions mediating a four-spin interaction in itinerant electronic square-net systems \cite{2021Hayami} and more recently magnetic dipolar interactions \cite{2022Paddison}.

Many centrosymmetric skyrmion hosts are intermetallic, and Gd-based, for instance \ce{GdPd2Si3} \cite{2019Kurumaji,2020Hirschberger} \ce{Gd3Ru4Al12} \cite{2019Hirschberger} and \ce{GdRu2Si2} \cite{2020Khanh, 2020Yasui}. More recently members of the \ce{Eu(Ga_{1-x}Al_x)4} series have been identified as candidate skyrmion hosts \cite{2022Moya}. For instance the end member of this series, \ce{EuAl4}, develops two different skLs under an applied magnetic field \cite{2022Takagi}. Furthermore, unique to the \ce{Eu(Ga_{1-x}Al_x)4} series for $x = 0.5$ and 1 is the development of a charge density wave (CDW) in zero field. 

\ce{EuGa2Al2}, the focus of this Letter, is electronically and structurally similar to \ce{EuAl4} and \ce{GdRu2Si2}; it is a rare-earth intermetallic that is expected to mediate magnetic exchange via long range RKKY interactions, the magnetic ions Gd$^{3+}$ and Eu$^{2+}$ are isoelectronic with spin only moments ($L = 0$, $J = 7/2$) \cite{2018Stavinoha, 2020Khanh, 2022Takagi}. Furthermore all three materials have been found to crystallise with $I4/mmm$ symmetry, where the magnetic ions form square nets in the $ab$ plane, which are not expected to support any geometrical frustration, and that are coupled along $c$ to form a three dimensional network of magnetic ions. The tetragonal symmetry of the crystal structure is thought to allow for the formation of multiple magnetic modulation vectors, that may then develop a double-Q square skyrmion lattice \cite{2020Khanh}. A topological hall effect analysis of the Hall resistivity in \ce{EuGa2Al2} suggests that a non-coplanar spin texture is stabilised when a magnetic field between 1.2 T $<$ H $<$ 1.8 T is applied parallel to the $c$ axis, and for temperatures below $\sim$ 7 K, hinting at the existence of a topologically non-trivial magnetic phase \cite{2022Moya}. 

Despite its electronic and structural similarity to \ce{EuAl4} and \ce{GdRu2Si2}, \ce{EuGa2Al2} orders with different magnetic and electronic structures in zero field \footnote{Private Communication} \cite{2021Kaneko, 2022Takagi}, indicating it has different underlying electronic interactions. \ce{EuGa2Al2} develops a CDW below T$_{\mathrm{CDW}}$ = 50 K \cite{2022Moya}, and three distinct magnetic phases below T$_1$ = 19.5 K, T$_2$ = 15 K and T$_3$ = 11 K, labelled the AFM1, AFM2 and AFM3 phases respectively \cite{2018Stavinoha, 2022Moya}. Building a theoretical model that can describe the formation of skyrmion lattices in the absence of DM exchange necessitates an accurate experimental determination of the electronic ordering phenomena in such systems and across their phase diagram. To this end, we performed neutron powder diffraction (NPD) and resonant x-ray scattering (RXS) experiments on high-quality single crystals of \ce{EuGa2Al2} that demonstrates the following. Firstly the onset of the CDW order broke the four-fold symmetry that was present in the $I4/mmm$ crystal structure, and stabilised orthorhombic domains with either $Immm(0,0,g)s00)$ or $Fmmm(0,0,g)s00)$ symmetry. Secondly the onset of the magnetic order below T$_1$ suppressed the CDW order, indicating the two electronic order parameters were in competition. Thirdly, we rigorously demonstrate that \ce{EuGa2Al2} formed a SDW with moments oriented perpendicular to the direction of the propagation vector in the AFM1 phase and a cycloid with moments in the $ab$ plane in the AFM3 phase, in contrast to what has been reported in the literature \cite{2022Moya}. 

In the $I4/mmm$ symmetry the Eu$^{2+}$ ions, Wyckoff position $2a$, form two-dimensional square layers in the $ab$ plane, and neighbouring Eu layers, which are separated along $c$, are translated by ($\frac{1}{2}$,$\frac{1}{2}$,$\frac{1}{2}$) owing to the I-centering that relates them. The Al and Ga ions sit between neighbouring Eu layers and are ordered across Wyckoff positions $4d$ and $4e$, where Al fully occupies the $4d$ site and Ga the $4e$ site \cite{2018Stavinoha}. Single crystals of \ce{EuGa2Al2} were grown in accordance with Ref. \citenum{2022Moya} and \citenum{2018Stavinoha}. RXS measurements were performed on the I16 beamline at Diamond Light Source \cite{2010Collins}. An as grown 1 $\times$ 1 $\times$ 0.5 mm$^3$ single crystal sample was fixed onto a Cu sample holder, and mounted onto a six-circle kappa goniometer. The ($00l$) was specular, and the ($0k0$) direction was used as an azimuthal reference. The incident energy was tuned to 6.97 keV, the Eu $L_3$ edge. Further details of the experimental set-up are given in the Sec. S1 of the Supplemental Material (SM).

\begin{figure}[ht]
\centering
\includegraphics[width =\linewidth]{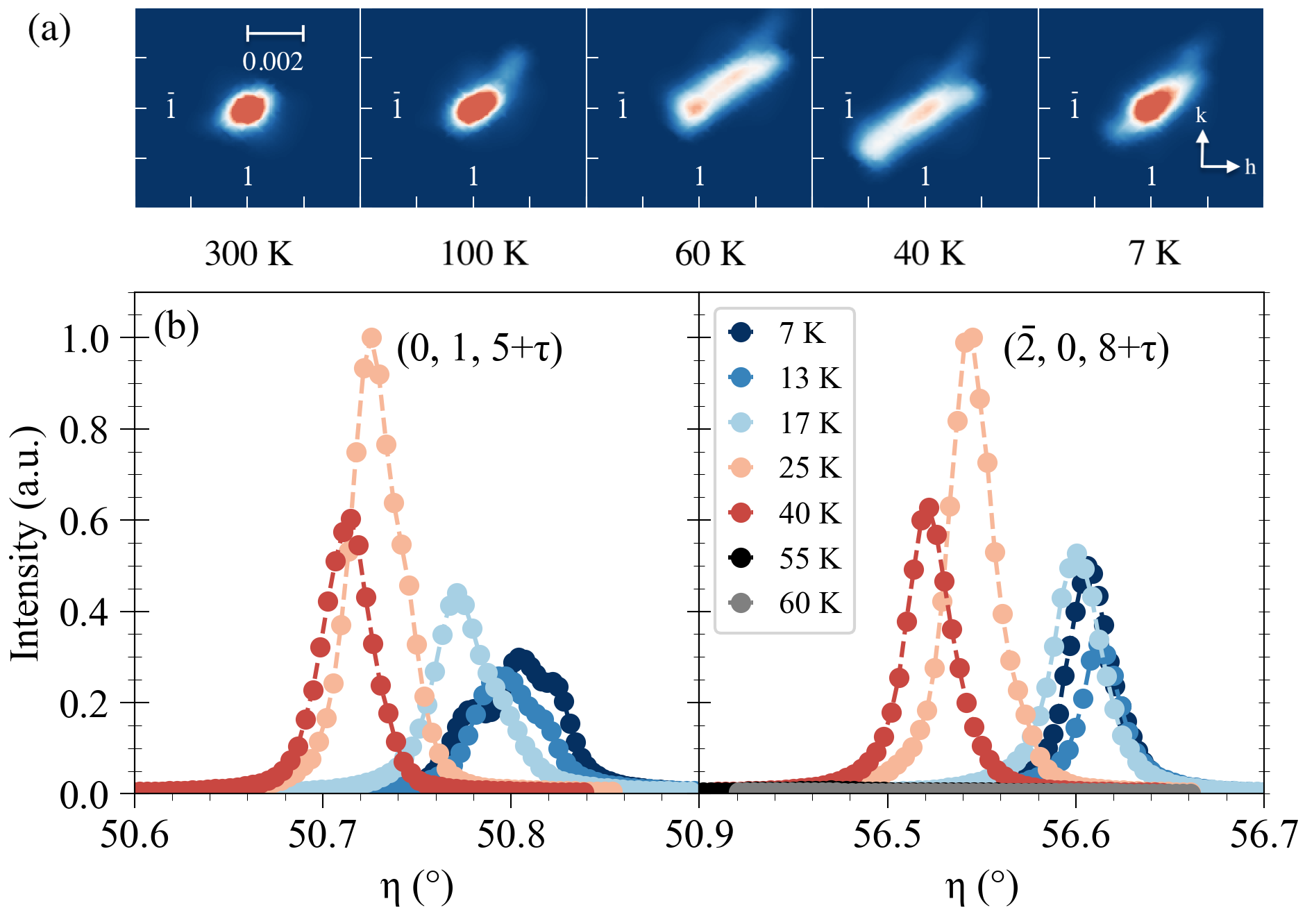}
\caption{\label{FIG::CDW} RXS data collected on (a) the (1,$\bar{1}$,8) charge reflection between 300 K and 7 K. (b) the (0,1,5+$\tau$) and ($\bar{2}$,0,8+$\tau$) CDW reflections between 7 K and 60 K. These data were collected using incident $\sigma$ polarised light.}
\end{figure}

At room temperature the crystal was indexed using the published $I4/mmm$ space group \cite{2018Stavinoha}. As the sample was cooled a number of distinct changes to the crystal structure were observed. Below 50 K satellite peaks appeared that were indexed with propagation vector \textbf{k} = (0,0,$\tau$), $\tau$ $\sim$ 0.125, which were identified to originate in a CDW, consistent with reports in the literature \cite{2022Moya}. Owing to the high reciprocal space resolution provided by the I16 beamline it was possible to observe for the first time subtle changes to the crystal structure as we cooled through the CDW and magnetic phase transitions. We observed an elongation of the structural Bragg reflections in the $hk$ plane below T$_{\mathrm{CDW}}$, indicating a loss to the four-fold symmetry that was present at 300 K. The possible subgroups of the displacive representation of the incommensurate CDW propagation vector, \textbf{k} = (0,0,$\tau$), consistent with a loss to the four-fold symmetry, are the orthorhombic space groups $Immm(0,0,g)s00$ and $Fmmm(0,0,g)s00$. Each of the two orthorhombic subgroups would produce a distinct splitting of (1,1,$l$) type reflections, as illustrated in Sec. S2 of the SM. For instance Fig. \ref{FIG::CDW}(a) shows the (1,$\bar{1}$,8), which begins to split close to the (1,1,0) direction as the temperature approaches T$_{\mathrm{CDW}}$, consistent with the presence of domains of $Immm(0,0,g)s00$ symmetry. Upon moving to different positions on the sample we observed a splitting of the (1,$\bar{1}$,8) that was consistent with the presence of both orthorhombic domains, Sec. S2 of the SM. This finding demonstrates that \ce{EuGa2Al2} is not a square-net magnet, similar to \ce{EuAl4} \cite{2022Ramakrishan}. Furthermore it implies the symmetry of the magnetically ordered phases cannot be higher than monoclinic and each orthorhombic domain could further split into two magnetic domains.

\begin{figure}[ht]
\centering
\includegraphics[width =\linewidth]{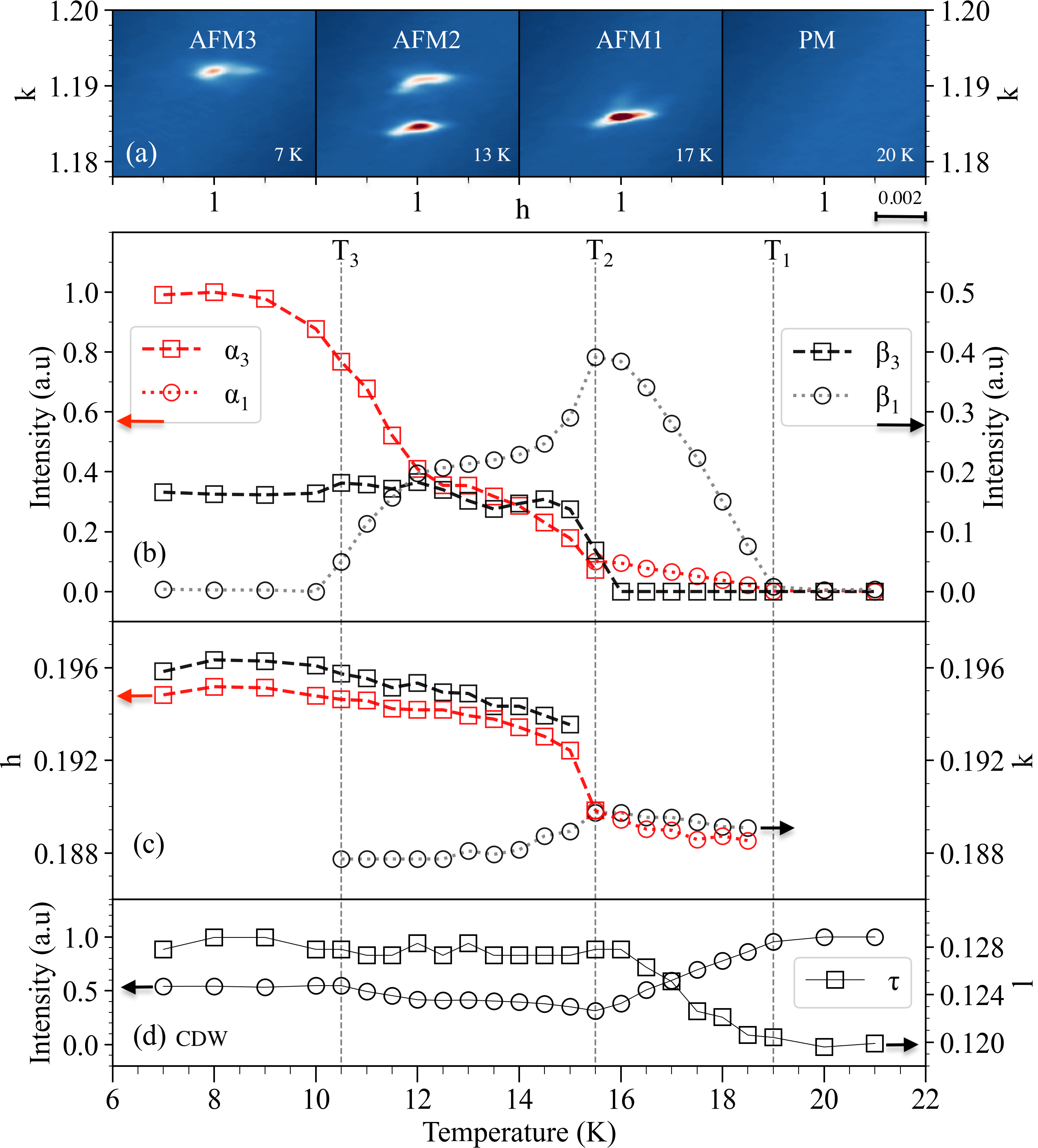}
\caption{\label{FIG::undistortedtempdep}(a) Projection of the (1,1+$\beta$,8) reflection in the $hk$ plane collected in the PM, AFM1, AFM2 and AFM3 phases. Temperature dependence of (b) the normalised integrated intensity and (c) propagation vectors of the magnetic satellites representative of the AFM3 phase, ($\alpha_3$,0,8) \& (0,$\beta_3$,8), and the AFM1 phase, ($\alpha_1$,0,8) \& (0,$\beta_1$,8). (d) Temperature dependence of the normalised integrated intensity and propagation vector of the (1,1,8+$\tau$) CDW reflection. These data were collected using incident $\sigma$ polarised light.}
\end{figure}

Below T$_1$ we measured several magnetic reflections that indexed with propagation vector, \textbf{k} = ($\alpha$,0,0) or \textbf{k} = (0,$\beta$,0), where $\alpha$ and $\beta$ varied between 0.196 and 0.188 depending on the temperature, Fig. \ref{FIG::undistortedtempdep}. Sudden changes to these satellite reflections appeared at T$_1$, T$_2$ and T$_3$, confirming the presence of the magnetically ordered phases previously reported in the literature \cite{2022Moya}, Fig. \ref{FIG::undistortedtempdep}(a). The $\alpha$ and $\beta$ satellites were observed at a position in the sample representative of a single orthorhombic domain, demonstrating that the orthorhombic domains, of which there can be four, were further split into two magnetic domains. No intensity was observed at reflections where $h+k+l=2n+1$, $n$ $\in$ $Z$, nor $\alpha$ and $\beta$ magnetic satellites centred about them, implying the I-centring relating the Eu1 and Eu2 ions was not broken in any of the three magnetic phases, as shown in Fig. S5 of the SM. 

We describe the temperature dependent changes observed for the ($\alpha$,0,8) and (0,$\beta$,8) magnetic satellites belonging to a single orthorhombic domain measured upon warming. For the ($\alpha$,0,8) reflection a single peak, labelled $\alpha_3$, was observed in the AFM3 phase, Fig. \ref{FIG::undistortedtempdep}(b). The intensity of this peak decreased to zero as the system was warmed to the AFM1 phase. At T$_2$, a second peak labelled $\alpha_1$ appeared, such that the $\alpha_1$ and $\alpha_3$ peaks coexisted over a temperature interval of 0.5 K. The intensity of the $\alpha_1$ peak was maximised at T$_2$ before steadily decreasing to zero as the system was warmed to the paramagnetic phase. A similar dependence was observed for (0,$\beta$,8), Fig. \ref{FIG::undistortedtempdep}(b), with the exception that the coexistence of the $\beta_1$ and $\beta_3$ peaks occurred over a wider temperature interval of 4 K. This suggests that the two peaks, $\alpha_1$/$\beta_1$ and $\alpha_3$/$\beta_3$ are representative of the AFM1 and AFM3 phases respectively, and originate in two competing magnetic phases.
Note that the difference in the coexistence region of the two magnetic domains can depend on microscopic characteristics such as their size. 

We also followed the evolution of the propagation vector and intensity of the (1,1,8+$\tau$) CDW reflection through the magnetic phase transitions to establish the coupling between these phenomena. The intensity of the CDW almost halved, while the value of $\tau$ steadily increased between T$_1$ and T$_2$, indicating that it was competing with the magnetic order. Note that the temperature dependence of over thirty different CDW reflections were collected as the crystal was warmed through the magnetically ordered phases, shown in Sec. S2 of the SM, which all showed the same behaviour. The electrons that give rise to the CDW are itinerant and expected to originate from the Al ions \cite{2016Kobata}, while the electrons responsible for the magnetic order are the localised 4$f$ electrons on the Eu sites. If the onset of the magnetic order polarises the itinerant electronic density that gives rise to the CDW, it is feasible that that it could destabilise the CDW order and thus cause its suppression. Indeed theoretically it has been shown that the spin of the conduction electrons tends to align with the underlying local moment texture in such itinerant magnets \cite{2016Batista}. Below T$_2$ we also observed a change to the relative intensity of different CDW reflections, Fig. \ref{FIG::CDW}(b), indicating the structure of the CDW is changing. We note that for \ce{EuAl4}, a CDW with orthorhombic symmetry with a similar propagation vector, \textbf{k} = (0, 0, 0.1781(3))) was observed, where the CDW order was not thought to originate in a simple nesting of the Fermi surface \cite{2016Kobata, 2022Ramakrishan}. 

Using magnetic symmetry analysis we identified nine different magnetic structures that the system could adopt, shown in Sec. S3 of the SM. The magnetic structures differ according to the moment direction adopted by the Eu ions, and also according to whether they are collinear (SDW) or non-collinear (helix and cycloid). As the resonant magnetic x-ray scattering (RMXS) cross section for the $\sigma\pi '$ channel is dependent on the dot product between the incident wavevector, $\hat{k}_i$, and the magnetic interaction vector the moment direction can be obtained by rotating the crystal relative to these vectors and measuring the scattered signal (azimuthal scan). 

\begin{figure}[ht]
\centering
\includegraphics[width =\linewidth]{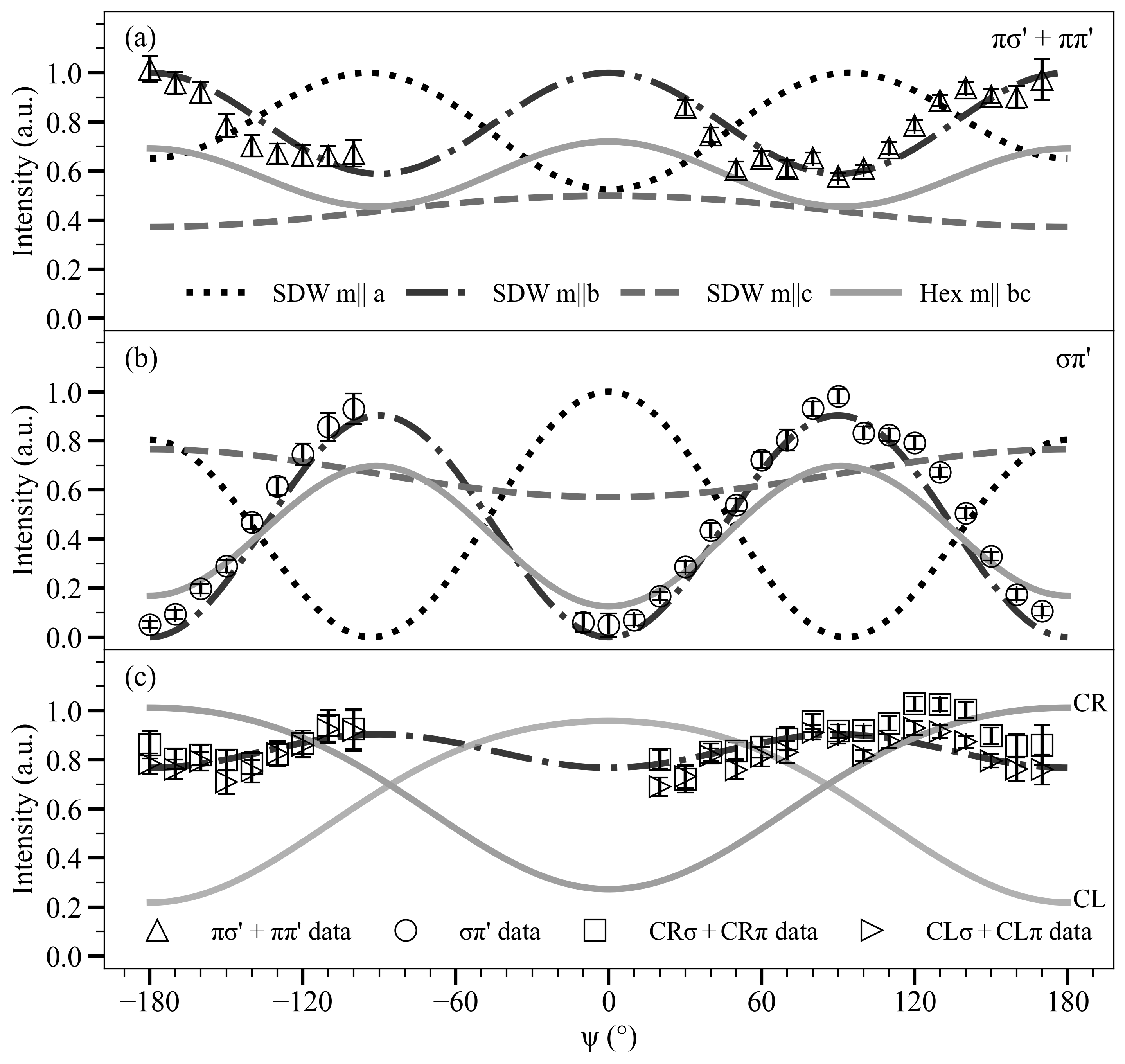}
\caption{\label{FIG::rawazi} Azimuthal scans collected in (a) $\pi\sigma '$ + $\pi\pi '$ channels (b) $\sigma \pi '$ channel and (c) the CR$\sigma$ + CR$\pi$ and CL$\sigma$ + CL$\pi$ channels for the (0,$\beta$,8) reflection. The lines represent the simulated azimuthal dependencies for the different magnetic structure solutions. Details of the simulations are presented in Sec. S4 of the SM. The unfilled markers represent the normalised data.}
\end{figure}

\begin{figure*}[ht]
\centering
\includegraphics[width =\linewidth]{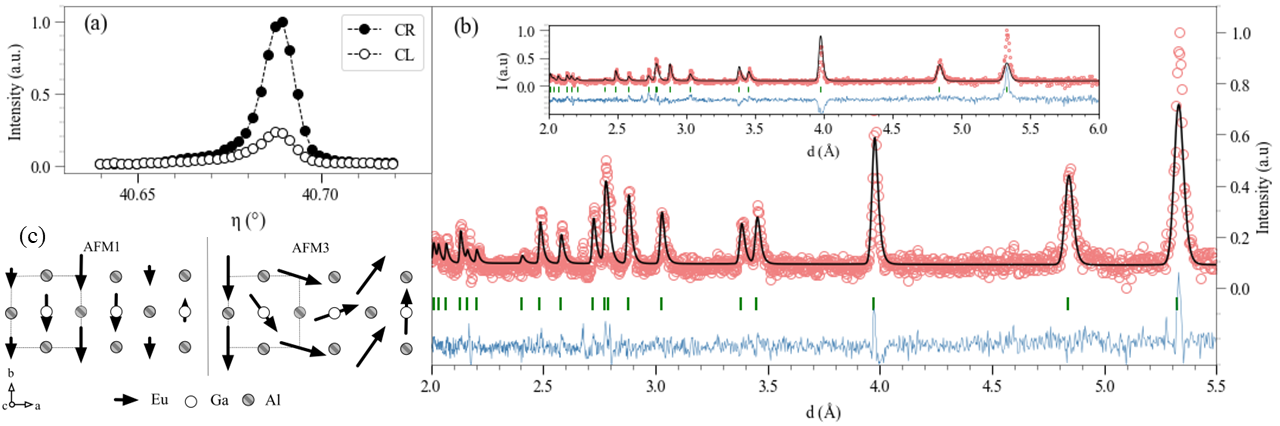}
\caption{\label{FIG::AFM3}(a) RMXS data collected on (0,$\beta$,8) magnetic satellite with CR and CL incident light. (b) Subtracted NPD data at 1.5 K (25 K data was subtracted) showing only magnetic reflections. Data is given by the unfilled red circles, the fit to the data using a circular cycloid with moments in the $ab$ plane by the solid black line, and their difference by the solid blue line. The green ticks mark the position of the magnetic reflections in d-spacing. The inset shows the subtracted data fit using a cycloid with moments in the $ac$ plane. (c) Experimentally determined magnetic structure of \ce{EuGa2Al2} in the AFM1 and AFM3 phases projected in the $ab$ plane. The black arrows represent the moments on the Eu ions, and the Ga and Al ions are denoted by the circles. The $I4/mmm$ crystallographic unit cell is shown by the solid black line.}
\end{figure*}

Fig. \ref{FIG::rawazi} shows the azimuthal scan collected on the (0,$\beta$,8) in the $\sigma\pi '$ channel at 17 K in the AFM1 phase. Maximums in the intensity of the scattered signal were measured at $\psi$ = 90$\degree$ and $\psi$ = -90$\degree$, when $\hat{k}_i$ was parallel to the $a$-axis, therefore a component of the magnetic moment was oriented along $a$. Furthermore as the diffracted intensity was zero at $\psi$ of 0, 180$\degree$, and -180$\degree$, a $b$ or $c$ axis component to the magnetic moment was not present. Measurements with incident circular light can be used to determine if the magnetic structure is non-collinear, as explained in Sec. S4 of the SM. The difference in the scattered intensity between the CR and CL light was zero across the majority of azimuthal values measured, Fig. \ref{FIG::rawazi}(c), indicating the AFM1 phase was collinear. The best fit to the azimuthal dependencies collected on the ($\pm\alpha$,0,8) and (0,$\pm \beta$,8) reflections measured with all four incident polarisation of light, details of which are given in Sec. S4 of the SM and shown for the (0,$\beta$,8) in Fig. \ref{FIG::rawazi}, was a SDW with moments aligned perpendicular to the direction of the propagation vector, consistent with the observations above. This magnetic structure solution transforms by a single irrep, which is consistent with the PM to AFM1 phase transition being second order in nature. The adoption of the moments perpendicular to the direction of the propagation vector is similar to that observed for \ce{EuAl4} \footnote{Private Communication} and \ce{Gd2PdSi3} \cite{2022Paddison}, suggesting an anisotropy term that may be common to many of the centrosymmetric skyrmion hosts. 

At 7 K, in the AFM3 phase, we found that the magnetic structure of the $\alpha$ and $\beta$ magnetic domains was non-collinear, owing to a difference in the magnetic intensity measured with CR and CL light, as shown for the (0,$\beta$,8) in Fig. \ref{FIG::AFM3}(a). A determination of moment direction using an azimuthal scan was not possible owing to the appearance of additional peaks that changed significantly with sample position, and hence azimuth. Given that a non-collinear magnetic structure would split each magnetic domain into a further two magnetic domains related by inversion symmetry, the appearance of the additional peaks was likely owing to the now quite complex domain structure. As such we employed the use of NPD to determine the ground state magnetic ordering. Several single crystal samples of \ce{EuGa2Al2} produced from the same growth were finely crushed to produce a 0.95 g polycrystalline sample, which was measured using the time-of-flight diffractometer WISH at ISIS \cite{2011Chapon}. The NPD data shown in Fig. \ref{FIG::AFM3}(c) was collected at 1.5 K, deep into the AFM3 phase, and a Rietveld refinement of the data conclusively showed that the magnetic structure is a cycloid with moments in the $ab$ plane. Details of the Rietveld refinement are given in Sec. S5 of the SM. The sudden changes to the propagation vector and intensity of the magnetic satellites at T$_2$, shown in Fig. \ref{FIG::undistortedtempdep}, is consistent with this phase transition being first order in nature. We question whether there are three distinct magnetically ordered phases, given that the AFM2 phase appears to be a coexistence of the AFM1 and AFM3 phases, Fig. \ref{FIG::undistortedtempdep}(a). Furthermore a region of phase coexistence is common following a first order phase transition. We suggest that the signature of the phase transition in the specific heat capacity at T$_3$ reported in Ref. \citenum{2018Stavinoha} may be related to the onset of a structural phase transition as indicated by small changes to the relative intensity of the CDW reflections observed between T$_2$ and T$_3$, Fig. \ref{FIG::CDW}, indicating a possible rearrangement of the CDW structure. While the change to the magnetic susceptibility at T$_3$, Ref. \citenum{2018Stavinoha}, may be related to a change in the magnetic domain pattern caused by a complete suppression of the AFM1 phase and the creation of inversion domains. We propose that if the CDW structure does transform below T$_2$ it may cause a change to the itinerant electronic density, which in turn may modify the RKKY interactions, proving a possible mechanism by which the cycloid m$||ab$ plane is stabilised.

In conclusion our study of \ce{EuGa2Al2} has shown that the onset of the CDW order breaks the four-fold symmetry present in the $I4/mmm$ crystal structure stabilising orthorhombic domains, which demonstrates \ce{EuGa2Al2} is not a square-net magnet. We find that the single crystal samples of \ce{EuGa2Al2} are composed of magnetic domains described by propagation vector ($\alpha$,0,0) or (0,$\beta$,0), which order in the AFM1 phase by a SDW with moments perpendicular to the direction of the propagation vector. We observed a suppression of the CDW order as the system ordered to form the SDW, suggesting the two electronic ordering phenomena were in competition, which in turn implies that the localised 4$f$ electrons and itinerant electronic density are coupled. Finally our results show that the ground state magnetic structure was a cycloid with moments in the $ab$ plane. Our findings map the zero-field magnetic and structural phases of \ce{EuGa2Al2}, revealing more complexity than was previously discovered, and demonstrating the requirement for high-resolution scattering studies to elucidate the true nature of the complex ordering present in such candidate centrosymmetric skyrmion hosts. 

\begin{acknowledgments}
The authors gratefully acknowledge the technical support from the team at the I16 Beamline, Diamond Light Source where the resonant x-ray scattering measurements were performed (Exp. MM30799-1).
\end{acknowledgments}
\bibliographystyle{apsrev4-2}

\end{document}